\documentclass[twocolumn,amsmath,amssymb,prb,aps]{revtex4}

\usepackage{graphicx}
\usepackage{dcolumn}
\usepackage{bm}

\begin{document}

\title{Room temperature electron spin coherence in telecom-wavelength quaternary quantum wells}

\author{W. H. Lau, V. Sih, N. P. Stern, R. C. Myers, D. A. Buell, A. C. Gossard, and D. D. Awschalom}
 \email{awsch@physics.ucsb.edu}

\affiliation{Center for Spintronics and Quantum Computation,
University of California, Santa Barbara, CA 93106}

\date{\today}
\begin{abstract}
Time-resolved Kerr rotation spectroscopy is used to monitor the
room temperature electron spin dynamics of optical
telecommunication wavelength AlInGaAs multiple quantum wells
lattice-matched to InP.  We found that electron spin coherence
times and effective g-factors vary as a function of aluminum
concentration. The measured electron spin coherence times of these
multiple quantum wells, with wavelengths ranging from $1.26$
$\mu$m to $1.53$ $\mu$m, reach approximately 100 ps at room
temperature, and the measured electron effective g-factors are in
the range from $-2.3$ to $-1.1$.
\end{abstract}
\maketitle

Recent developments in semiconductor-based spintronics and quantum
computation~\cite{Wolf:2001,Awschalom:2002} have generated intense
interest in examining the spin dynamics in a wide range of
semiconductor material systems at telecom
wavelengths~\cite{Tackeuchi:2005,Hall:1999,Akasaka:2004,Marshall:2002}
due to their potential applications in spin-dependent optical
modulators and switches for optical
telecommunications~\cite{Marshall:2002,Tackeuchi:1990,Nishikawa:1995,Oestreich:1996}.
The optical wavelengths of $1.31$ ${\mu}$m and $1.55$ ${\mu}$m
allow a minimum dispersion and signal loss in standard
silica-based fiber-optic networks. Spin-dependent all-optical
switching in semiconductor quantum well
etalons~\cite{Nishikawa:1995} and spin-dependent ultrafast optical
gain modulation in microcavity lasers have been recently
demonstrated~\cite{Oestreich:1996}. In addition, differential
quantum efficiency as high as $64\%$ is attainable for
all-epitaxially grown telecom-wavelength vertical cavity surface
emitting lasers (VCSELs) with InP lattice-matched AlInGaAs alloys
for quantum well active regions and AlGaAsSb alloys for highly
reflective distributed Bragg reflectors operating above room
temperature~\cite{Buell:2006,Feezell:2005}. The possibility of
integrating laser sources, optical modulators, and optical
switches on a single photonic integrated chip with added
functionalities based on spin degrees of freedom makes AlInGaAs
alloys ideal candidates for spintronics.

In this letter, we present room temperature measurements of the
electron spin coherence times and effective g-factors in AlInGaAs
multiple quantum wells, which are optimized active regions of
VCSELs with wavelengths in the range $1.26-1.53$ $\mu$m designed
for optical telecommunication
applications~\cite{Buell:2006,Feezell:2005}. We characterize the
samples using photoluminescence and measure the electron spin
dynamics using pump-probe optical spectroscopy. The electron
effective g-factors and spin coherence times are found to vary
with aluminum alloy concentration from $-2.3$ to $-1.1$ and from
$50$ ps to $100$ ps, respectively.

\begin{figure}\includegraphics{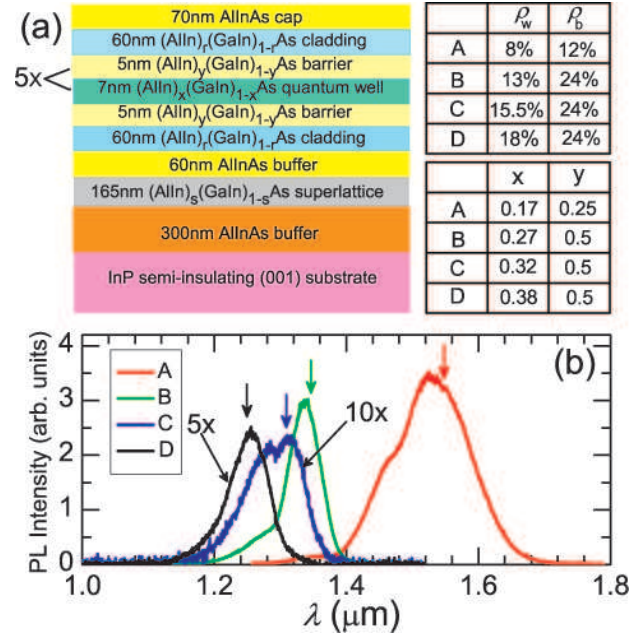}\caption{\label{fig1}
(Color)  (a) Sample structure of multiple quantum wells, where $r$
$=$ $0.75$ and $s$ $=$ $0.82$, and $5$x signifies the growth
repetition. The average aluminum concentration of the quantum
wells and barriers are $\rho_\text{w}$ and $\rho_\text{b}$,
respectively. (b) Photoluminescence spectra at room temperature.
The vertical arrows indicate the positions of the calculated
effective band gaps. The intensity of the PL spectra for Sample C
and Sample D is scaled for clarity.}
\end{figure}

Our samples are comprised of a set of quaternary alloy AlInGaAs
multiple quantum wells grown using digital alloy molecular beam
epitaxy. These samples are composed of digital alloys of ternaries
(Ga$_{0.47}$In$_{0.53}$As and Al$_{0.48}$In$_{0.52}$As, referred
to as GaInAs and AlInAs, respectively) lattice-matched to
InP~\cite{Buell:2006}. The sample structure is of the form:
AlInAs/(AlIn)$_{r}$(GaIn)$_{1-r}$As/5x[(AlIn)$_{y}$(GaIn)$_{1-y}$As
/(AlIn)$_{x}$(GaIn)$_{1-x}$As]/(AlIn)$_{y}$(GaIn)$_{1-y}$As/(AlIn)$_{r}$
(GaIn)$_{1-r}$As/AlInAs/(AlIn)$_{s}$(GaIn)$_{1-s}$As/AlInAs/InP
substrate, shown schematically in Fig.1(a). All samples are
undoped. The photoluminescence (PL) spectra of these samples are
excited non-resonantly with $2.5$ mW of $800$ nm light from a
tunable mode-locked Ti:sapphire laser and measured using a liquid
nitrogen cooled InGaAs photodiode array detector [Fig. 1(b)]. The
PL peaks of the samples at room temperature occur at $1.53$ $\mu$m
(Sample A), $1.34$ $\mu$m (Sample B), $1.30$ $\mu$m (Sample C),
and $1.26$ $\mu$m (Sample D), and the corresponding energies are
$0.811$ eV, $0.925$ eV, $0.954$ eV, and $0.984$ eV, respectively.
The PL intensity decreases with increasing aluminum concentration
$\rho_\text{w}$ due to Al incorporated impurities, such as oxygen,
and lower confinement of the electron wave functions arising from
lower conduction band offsets. The PL spectrum of Sample C
consists of two broad peaks ($1.28$ $\mu$m and $1.30$ $\mu$m), and
the $1.28$ $\mu$m peak might be due to impurity broadening. We
calculate the effective band gaps of the quantum wells using a
generalized fourteen-band ${\bf K \cdot p}$ envelope-function
theory~\cite{Lau:2005,Lau:2006}, and the calculated effective band
gaps for Samples A-D are $1.55$ $\mu$m, $1.35$ $\mu$m, $1.30$
$\mu$m, and $1.25$ $\mu$m, respectively.

\begin{figure}\includegraphics{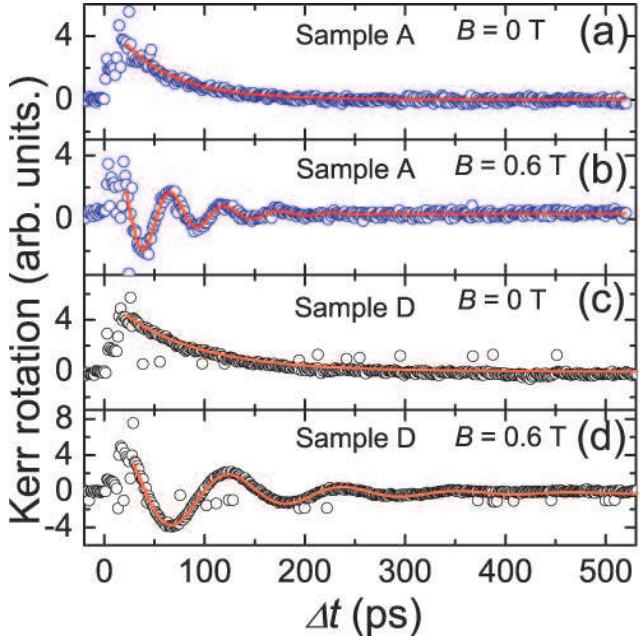}\caption{\label{fig2}
(Color) Time-resolved Kerr rotation of Sample A at room
temperature for (a) $B$ $=$ $0$ T, and (b) $B$ $=$ $0.6$ T, and
Sample D for (c) $B$ $=$ $0$ T, and (d) $B$ $=$ $0.6$ T. Offset is
subtracted for clarity. Circles are measurements and lines are
fits using the expression for $\theta(\Delta t)$.}
\end{figure}

Time-resolved Kerr rotation (TRKR), an optical pump-probe
spectroscopic technique~\cite{Crooker:1997}, is used to probe the
electron spin dynamics. An optical parametric amplifier pumped by
a regeneratively amplified Ti:sapphire mode-locked laser produces
$\sim200$ fs duration pump and probe pulses tunable from $1.1$
$\mu$m to $1.65$ $\mu$m with a repetition rate of $100$ kHz, whose
relative delay is adjusted by a mechanical delay line. The
helicity of the pump beam for spin injection is modulated with a
photoelastic modulator at $42$ kHz, and the linearly polarized
probe beam for spin detection is mechanically chopped at a
frequency of $800$ Hz. The pump and probe beams are focused to a
spot size of $\sim50\mu$m on the sample, which is mounted between
the two poles of an electro-magnet which provides a magnetic field
$B$ up to $0.6$ T and perpendicular to the direction of the pump
and probe beams. The circularly polarized pump beam excites
electron and hole spins which precess at the Larmor frequency
around the transverse applied magnetic field. The typical pump and
probe powers used in the measurements are $1$ mW and $0.1$ mW,
respectively. The rotation of the linear polarization axis of the
probe beam is proportional to the net electron spin polarization
along the probe's normal incidence. Using a balanced photodiode
bridge combined with a lock-in detection technique, the electron
spin precession is measured with subpicosecond temporal
resolution~\cite{Crooker:1997}.

\begin{figure}\includegraphics{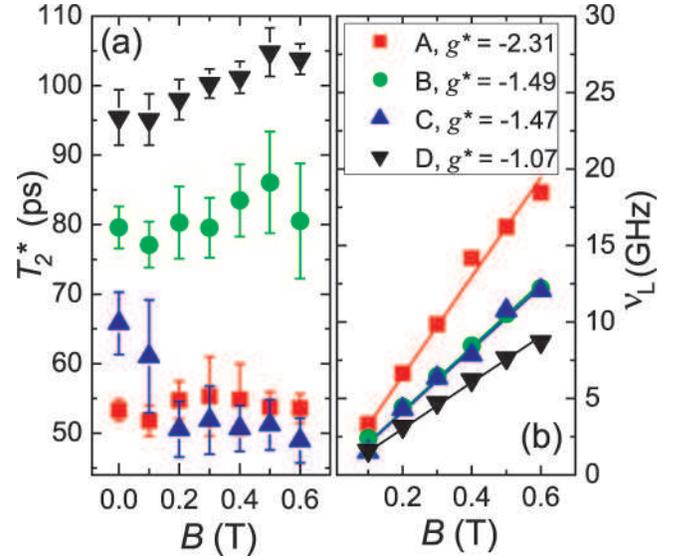}\caption{\label{fig3}
(Color) (a) Electron spin coherence time $T_2^\ast$, and (b) spin
precession frequency $\nu_{\text{L}}$ as a function of magnetic
field $B$ at room temperature. The corresponding lines are linear
fits.}
\end{figure}

In Fig. 2, we show representative TRKR scans for Sample A ($8\%$
Al) and Sample D ($18\%$ Al) at room temperature with applied
magnetic fields $B$ $=$ $0$ T and $B$ $=$ $0.6$ T. The electron
spin magnetization precesses in the plane perpendicular to the
applied magnetic field, and the Kerr rotation angle as a function
of time delay $\Delta t$ can be described by an exponentially
decaying cosine: $\theta(\Delta t)$ $=$
$\theta_\text{o}\exp(-t/T^\ast_2)\cos(2 \pi \nu_\text{L} \Delta
t)$, where $\theta_\text{o}$ is the initial amplitude of the
electron spin polarization, $\nu_\text{L}$ is the Larmor
precession frequency $\nu_\text{L}$ $=$ $\text{g}^\ast
\mu_\text{B} B/h$, where $\text{g}^\ast$ is the effective in-plane
electron g-factor of electrons in the quantum wells,
$\mu_\text{B}$ is the Bohr magneton, $h$ is Planck's constant, and
$T^{\ast}_2$ is the electron spin coherence time. The data are fit
by adjusting the parameters $\theta_\text{o}$, $T^{\ast}_2$, and
$\text{g}^\ast$. The lack of additional frequency components and
the agreement of $\text{g}^\ast$ with the expected
value~\cite{Lin:2004,Madelung:2003} for electrons indicate that
hole spin precession is not observable.

\begin{figure}\includegraphics{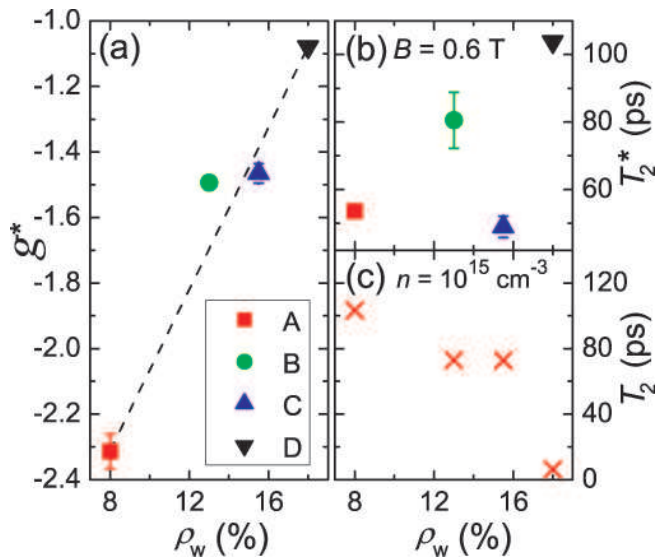}\caption{\label{fig4}
(Color) Electron effective g-factor $\text{g}^\ast$ (a), electron
spin coherence time $T_2^\ast$ for $B$ $=$ $0.6$ T (b), and
calculated $T_2$ (c) as a function of $\rho_\text{w}$. Red crosses
represent the calculated $T_2$ assuming $\mu$ $=$ $2000$
cm$^2$/Vs. $n$ is electron density used in the calculations. Line
is guide to the eyes.}
\end{figure}

The magnetic field dependence of $T^{\ast}_2$ at room temperature
for all four samples is plotted in Fig. 3(a).  All the samples
exhibit weak magnetic field dependence. The corresponding Larmor
frequencies $\nu_\text{L}$ of all samples have a linear magnetic
field dependence [Fig. 3(b)], yielding
$\left|\text{g}^\ast\right|$ $=$ $2.31$ (Sample A),
$\left|\text{g}^\ast\right|$ $=$ $1.49$ (Sample B),
$\left|\text{g}^\ast\right|$ $=$ $1.47$ (Sample C), and
$\left|\text{g}^\ast\right|$ $=$ $1.07$ (Sample D). Although the
sign of $\text{g}^\ast$ cannot be determined from our
measurements, the fact that bulk InAs has a large and negative
g-factor, while bulk AlAs has a small positive g-factor indicates
that $\text{g}^\ast$ is negative~\cite{Lin:2004,Madelung:2003}.
The electron effective g-factor as a function of $\rho_\text{w}$
at room temperature is plotted in Fig. 4(a), showing
$\text{g}^\ast$ increases with increasing $\rho_\text{w}$
consistent with the expected trends~\cite{Lin:2004,Madelung:2003}.

The electron spin coherence time $T^\ast_2$ as a function of
$\rho_\text{w}$ at room temperature with applied magnetic field
$B$ $=$ $0.6$ T is displayed in Fig. 4(b). The measured electron
spin coherence times are in the range from $50$ ps to $100$ ps.
$T^\ast_2$ increases with increasing $\rho_\text{w}$ except for
Sample C. The observed nonlinear $\rho_\text{w}$ dependence of
$T^\ast_2$ is due to the combined effects of orbital scattering
and internal effective magnetic field in the quantum
wells~\cite{Lau:2005,Lau:2006}.

Electron spin relaxation near room temperature is dominated by the
D'yakonov-Perel' mechanism~\cite{Dyakanov:1971,Lau:2006}, and the
spin relaxation rate is $T^{-1}_2$ $\propto$ $\Omega^2
\tau_\text{o}$, where $\Omega$ is the momentum-dependent
precession frequency about the internal effective magnetic field
of the quantum wells and $\tau_\text{o}$ is the orbital coherence
time, which is proportional to the electron mobility $\mu$. As
$\rho_\text{w}$ increases, $\Omega$ increases due to band
structure effects on the effective internal magnetic field,
causing $T_2$ to decrease with increasing $\rho_\text{w}$. To
illustrate these effects, we calculate the spin coherence times
$T_2$ as a function of $\rho_\text{w}$ for Samples A-D assuming a
constant mobility for all samples using a generalized $14$-band
${\bf K \cdot p}$ envelope-function
theory~\cite{Lau:2005,Lau:2006,parameters}, and the results are
shown in Fig. 4(c). However, the observed trends in $T^\ast_2$
might be partly due to the nonlinear $\rho_\text{w}$ dependence of
the electron mobility in these samples. As $\rho_\text{w}$
increase, $\tau_\text{o}$ decreases due to orbital scattering,
causing $T_2$ to increase with increasing $\rho_\text{w}$.
Finally, imperfection in crystal growth may lead to structural
inversion asymmetry in the quantum wells, which generates an
additional internal effective magnetic field contributing to
electron spin decoherence~\cite{Lau:2005}.

We have measured the electron spin coherence times and electron
spin precession frequencies in telecom-wavelength quaternary
multiple quantum wells and observed the electron spin coherence
times of $\sim100$ ps at room temperature. Our measurements reveal
that the electron effective g-factor can be tuned by varying the
aluminum alloy concentration. These results demonstrate the
possibility of optical modulation using spin manipulation in this
material system.

We wish to acknowledge the support of Intel and MARCO. N. P. Stern
acknowledges the support of the Hertz Foundation.


\end{document}